\documentclass[sigconf]{acmart}

\usepackage{acronym}

\AtBeginDocument{%
  \providecommand\BibTeX{{%
    \normalfont B\kern-0.5em{\scshape i\kern-0.25em b}\kern-0.8em\TeX}}}

\acmConference[CHIIR '24]{Conference on Human Information
Interaction and Retrieval}{March 10--14, 2024}{Sheffield, UK}

\begin{abstract}
While the topic of listening context is widely studied in the literature of music recommender systems, the integration of regular user behavior is often omitted.
In this paper, we propose PACE (PAttern-based user Consumption Embedding), a framework for building user embeddings that takes advantage of periodic listening behaviors.
PACE leverages users' multichannel time-series consumption patterns to build understandable user vectors.
We believe the embeddings learned with PACE unveil much about the repetitive nature of user listening dynamics. 
By applying this framework on long-term user histories, we evaluate the embeddings through a predictive task of activities performed while listening to music. 
The validation task's interest is two-fold, while it shows the relevance of our approach, it also offers an insightful way of understanding users' musical consumption habits. 

\end{abstract}

\copyrightyear{2024}
\acmYear{2024}
\setcopyright{acmlicensed}\acmConference[CHIIR '24]{Proceedings of the 2024 ACM SIGIR Conference on Human Information Interaction and Retrieval}{March 10--14, 2024}{Sheffield, United Kingdom}
\acmBooktitle{Proceedings of the 2024 ACM SIGIR Conference on Human Information Interaction and Retrieval (CHIIR '24), March 10--14, 2024, Sheffield, United Kingdom}
\acmDOI{10.1145/3627508.3638299}
\acmISBN{979-8-4007-0434-5/24/03}

\begin{document}

\title{Modeling Activity-Driven Music Listening with PACE}

\author{Lilian Marey}
\email{mareylilian@gmail.com}
\affiliation{%
    \institution{Deezer Research - LTCI, Télécom Paris}
    \city{Paris}
    \country{France}
}

\author{Bruno Sguerra}
\email{research@deezer.com}
\affiliation{%
    \institution{Deezer Research}
    \city{Paris}
    \country{France}
}
\author{Manuel Moussallam}
\email{research@deezer.com}
\affiliation{%
    \institution{Deezer Research}
    \city{Paris}
    \country{France}
}




 \maketitle

\section{Introduction}

The Oxford English Dictionary defines taste as ``The sense of what is appropriate, harmonious, or beautiful; the discernment and appreciation of the beautiful in nature or art''. Traditionally, musical taste is often declarative. Answers to questions such as ``Do you like the music of this artist?'' or ``What do you think of this kind of music?'' are considered to give a general idea of a person's taste. 
However, people's taste for art is shaped by the technological context, in particular by the accessibility of the medium in question. 
In recent years, most music listening has been done online, through streaming platforms.
This change has profoundly impacted people's relationship with music \cite{datta2018changing, fuentes2019soundtracking} for at least two reasons: first, the simplicity of listening (potentially everywhere and at any time), and second, the vast range of tracks now available at low cost.
In order to offer their users personalized recommendations, these services make the underlying assumption that interactions between the user and the catalog serve as accurate and exhaustive traces of musical taste. In fact, these interaction logs, collected during the use of streaming services, are precious data insofar as they can reveal behaviors listeners themselves are sometimes not aware of, or which they would not declare when asked during discussions. 
However, interactions between users and the catalog are shaped by different modalities of music consumption. For example, the practice of soundtracking (listening to music to accompany an activity) is a relatively recent one and is becoming more widespread \cite{fuentes2019soundtracking}. When soundtracking, users show a preference for songs that differ from their usual, e.g., a student might listen to lofi music to accompany their studying sessions \cite{wang2020lofi}. In this situation, the student employs music as a means of helping them focus, so the consumed music might be different from the one the student listens to for leisure or with friends. We call those streams \textbf{activity-driven listening}. In these situations, the choice to listen to music in the first place, and the consequent listened content, are highly dependent on the activity performed \cite{fuentes2019soundtracking, north2004uses}. 

Given the widespread use of soundtracking, distinguishing the logs generated by this practice from others is crucial for enhancing user taste models and refining recommendation methods. Being closely associated with specific activities, soundtracking allows us to leverage a fundamental aspect of human behavior: the inclination to follow recurring patterns. For instance, individuals commonly dedicate weekdays to work-related tasks, reserving weekends for leisure and relaxation \cite{ginoux2020did, north2004uses}. This inherent pattern facilitates the extraction of listening trends driven by users' activities from the recorded logs. In this paper, thus, we focus on the notion of listening regularity. The hypothesis we want to challenge is the following: can patterns in users' logs be leveraged for characterizing activity-driven listening?

% Since we are far from fully understanding human behavior in its entirety, mathematical language allows us to model it via embedding techniques, i.e. representing individuals in a highly simplified form. That said, each technique has its own specificities and tradeoffs: what component of human behavior one wants to highlight the most? In our case, the objective is to capture regularity in music listening behavior.
We therefore propose PACE (PAttern-based user Consumption Embedding), a new method for embedding user consumption histories, based on the encoding of regular listening behaviors. 
PACE leverages common weekly patterns of consumption that are used to generate user embeddings, characterizing the different consumption uses found in music streaming services. 
As some soundtracking activities are particularly regular, one way of evaluating our embeddings is to access if they capture activity-driven behaviors. Thus, we evaluate PACE through an activity prediction task. This step was made possible thanks to a survey carried out among Deezer (a major music streaming service) customers, which asks users what activities they use to practice while listening to music.

To summarize, our paper includes two major contributions: 

 \textbf{(1)}: the introduction of PACE, a framework for encoding users' histories, specializing embeddings on regular consumption patterns;

 \textbf{(2)}: the design of predictive models that shed light on the intrinsic relationships between regular listening behavior and the practice of soundtracking.

\section{Related work}

In order to answer the question  ``what determines the choice to listen to music?'', one might refer to the idea of intention. According to \cite{north2004uses, Volokhin_Agichtein_2018}, listeners may choose what they listen to according to the expected effect the music will have on their psyche (relaxation, concentration, motivation, etc.). The notion of intention is perhaps the most fundamental, insofar as it stems from psychological impulses. However, these intentions can be conditioned by many external factors of the user experience, such as the time of day, weather, location, or an activity being performed. This observation paves the way for the most widespread approaches in the music consumption literature, all grouped together in the so-called study of \textbf{listening context}. 

In \cite{adomavicius2010context}, Adomavicius et al. establish a first taxonomy of the general field of \ac{CARS}, classifying on the one hand the different types of contextual knowledge according to their observability (e.g., the listening timestamp is often an observable context, whereas the user's mood is in general unobservable) and according to the static/dynamic dimension, as well as the ways of integrating context in algorithms. In the literature, a number of variables are used to model the context of listening events. In \cite{yang2023nested}, the authors build a framework where several classes of contextual factors are nested (information retrieval, situational, personal, and social and cultural contexts). Accordingly, listening context can be inferred by a large set of features that characterize the user/item interactions.
In \cite{hansen2020contextual}, C. Hansen et al. illustrate the influence of time and device context in the embedding of listening sessions. In a similar approach, \cite{braunhofer2013location, cheng2016effective} use listener location, and \cite{sen2015sensors} uses sensor data, as weather or light levels. These contextual variables are observable and so relatively easy to collect, which is why they are present in many context-aware approaches. Beyond employing accessible data, several methods have been proposed to capture more sophisticated contexts: \cite{wang2018context} predicts user emotion, and \cite{oh2022implicit} infers context in a latent space without predefined labels.

These works generally deal with user histories over short periods of time and do not benefit from a long-term approach. Also, in the methods described above, user representations are a collection of successive snapshots containing contextual information, but the choice of time windows are often quite coarse, and a full temporal comprehensive profile of the user's behavior is not created. This is a shortfall, as it fails to take into account the regular aspects of human activity with precision. 
As we show over the next sections, our approach, on the other hand, highlights temporal hourly weekly consumption patterns that we might be unaware of, as it has rarely been studied in the literature.

For activity-driven listening, the activity practiced can of course be seen as part of the context. Some recent works seek to predict activity, but mainly with a content activity-tagging approach \cite{dias2014user, ibrahim2020audio, ibrahim2022exploiting}. On the latter, although some tracks are particularly well known for being appropriate in certain contexts (e.g., \textit{Eye of the Tiger} by Survivor for doing sports), it seems quite likely that the association between tracks and activities is highly user-dependent (some people will prefer calm rather than energetic music to wake up).
This is why this paper focuses on a user modeling-based approach.

\section{Methodology}
In PACE, each user's consumption history is encoded as a multivariate time series, i.e. time series composed of several dimensions, called channels. 
Each channel captures specific insights from the user's history, as detailed in the next section. 
The choice of these channels is shaped by our knowledge regarding common user consumption aspects. 
From these time series, we derive a dictionary containing a fixed number of stereotyped behaviors, or ``atoms''. 

While detecting the atoms, user's historical consumption time series are projected onto the dictionary to generate user embeddings. Unlike conventional user embedding techniques like matrix factorization, where users are represented in a latent space, each component of our vectors directly corresponds to an atom, which is, essentially, a time series. These time series encode distinct listening patterns on a weekly scale across various channels, surfacing information to help understand user behavior. Since the atoms' channels are understandable, interpreting these embeddings is doable given common weekly activity patterns and knowledge on music consumption modes from the literature.

\subsection{Encoding Consumption Histories}
The collected user's stream histories are issued from Deezer, and cover a period of a year and a half (January 2022 to May 2023). 
In the collected data, for each user, every music listening event has been recorded. 
For each event, we have access to several pieces of information, such as the listening timestamp, the song identifier, and the origin of the stream (if the user accessed the song by themselves, or it came from algorithmic recommendation).

\begin{figure}[ht]
  \centering
  \caption{Aggregation process for a particular user: stream count over the whole dataset (top), weekly scale aggregation (middle), after convolution and normalization (bottom).
}
  \includegraphics[width=.45\textwidth]{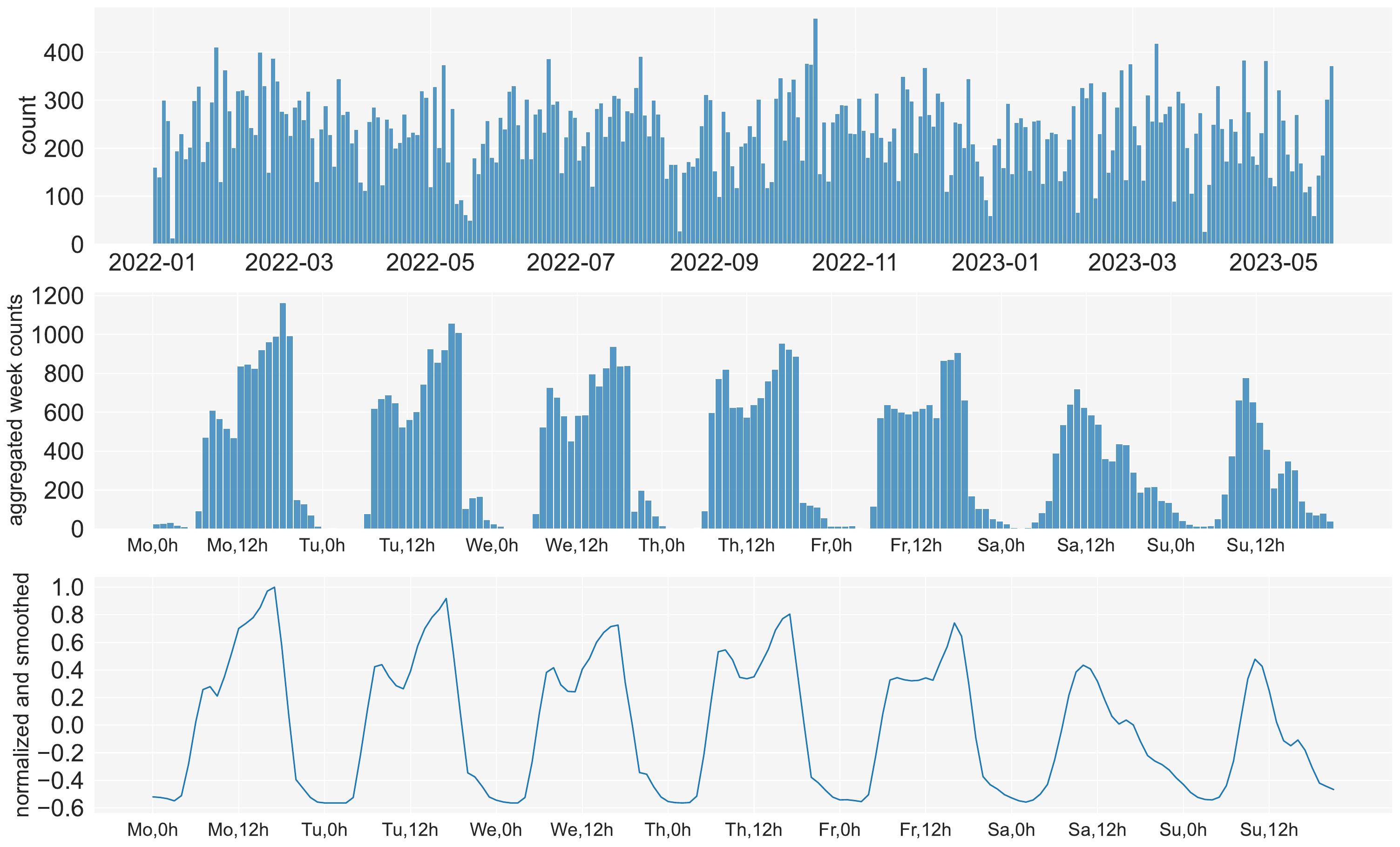}
  \label{fig:img1}
\end{figure}

Our goal is to capture temporal regularities in user behavior, and as \cite{taylor2018forecasting} shows, the most important seasonal effects in business time series are in general at yearly and weekly scales. Given the length of our data, we focus on weekly cycles. Therefore, the built signals are based on counts of subsamples of streams, over one-hour windows, and they are aggregated on a weekly hourly scale (e.g., a value is assigned to Tuesdays from 10 to 11 a.m.). 
This aggregation allows regular activity patterns to reveal themselves (one example is shown in Figure~\ref{fig:img1}).

Since the collected listening logs contain a wide range of behavioral information, we encode histories on several channels to better represent different user consumption patterns: 

\noindent \textbf{Volume}: the first channel we choose is a simple count of the number of streams. This highlights periods of regular listening, and encodes the user's general weekly behavior.
    
\noindent \textbf{Repetition}: a whole literature focuses on repeat consumption~\cite{benson2016modeling, anderson2014dynamics}. For example, in~\cite{sguerra2022discovery, sguerra2023ex2vec} authors show the phenomenon of the \textit{Mere Exposure Effect}, where multiple repetitions affect the evolution of users' interest, for example by increasing the temporal gaps between two successive consumption of the same song, over the repetitions. 

In this context, we use repeat behavior to gain insight into a user's intent when listening to music. On the one hand, choosing new content signifies an intent to discover, perhaps exploring songs by an artist or within a preferred genre. On the other hand, opting for known songs implies an awareness of the emotional impact they can have. Repeated listening indicates a stronger intent to control music's influence on one's mood. Therefore, we define a channel as a measurement for the repetitiveness of listening behavior.

\noindent \textbf{Organicity}: the music consumption in streaming platforms have different ``origins'', for instance, users might look for songs and albums themselves from their library or from the search bar, or enjoy automatic algorithmic music recommendation. 
This distinguishes \textit{organic} streams from \textit{algorithmic} ones.
In an organic context, the user makes a choice about the music listened to, and is in a decision-making position with regard to the content consumed, while in a recommendation driven session, the user can trade some of the control they have for discoverability or convenience.
From the considered listening histories, about 80\% of the streams come from organic usage. 
However, this rate is subject to variations among users and time of the week, which is why we build a channel to account for it.

\noindent \textbf{Liked}: furthermore, \cite{north2004uses} shows that the music chosen in collective listening contexts differs strongly from the user's fundamental taste (approaching a collective taste).
As a proxy for this information, we account for streams of users' liked content, i.e. songs the user assigned as favorite tracks, or songs from albums favorited by the user.
Therefore, a measure of the alignment of the music listened to with the user's more precise fundamental taste can be computed. \\

For each user, we build a multivariate time series, with the $4$ channels described above, each with a length of 168 (number of hours in a week). 
We define user signals tensor $\textbf{S}$, such that for a user $u$, a channel $c$, and a weekly $1$-hour window $t$:
\begin{equation*}    
    \textbf{S}_{u, c, t} = \frac{1}{|\{a \in A, a \sim t \}|}\sum_{a \in A, a \sim t}F_c(u, a),
\end{equation*}
where $\{a \in A, a \sim t \}$ is the set of all the $1$-hour windows of the database corresponding to their weekly equivalent $t$, and $F_c$ is the function coding channel $c$, defined as follows: $F_{\textbf{volume}}(u, a)$ counts the number of items listened by $u$ in $a$, and $F_{\textbf{repetition}}, F_{\textbf{organicity}}, F_{\textbf{liked}}$ respectively compute the ratio in $a$ of streams of music the user listened more than $3$ times overall, streams tagged \textit{organic}, and streams liked by the user. 
Note that streams listened to for less than 30 seconds are not considered.

In order to fade the strict $1$-hour delimitation, a convolution by a constant filter of length $3$ is applied to each time series channel.
In addition, to focus on finding patterns in these time series, we normalize the series on each channel for each user, fixing its mean to $0$ and its maximum absolute value to $1$.

\subsection{Detecting Listening Behavior Patterns}
To highlight typical behaviors, we use dictionary learning \cite{mairal2009online}. 
This allows the detection of a fixed number of time series (called atoms), that are computed as minimizers of a signal reconstruction error combined with a sparsity constraint.

As our data are multichannel series, we use the implementation of multivariate dictionary learning given by \cite{dupre2018multivariate}.
This implies learning multivariate atoms, stored in a tensor dictionary $\textbf{D}$. Thus, the optimization problem we tackle is the following:
\begin{center}
$\hat{\gamma} , \hat{\textbf{D}} = argmin_{\gamma, \textbf{D}}\sum_{c\in C}||\textbf{S}^c -  \textbf{D}^c\gamma||_F^2 + \lambda ||\gamma||_1$,
\end{center}
where $C$ is the set of channels, $\textbf{S}^c$ and $\textbf{D}^c$ are matrix extracted on channel $c$ from $\textbf{S}$ and  $\textbf{D}$, $\gamma$ is users' \textit{codes}, $||.||_F$ is the Frobenius matrix norm, and $\lambda$ is the regularization parameter.

\begin{figure}[ht]
  \centering
  \caption{Creating a specific user's embedding.}
  \includegraphics[width=.45\textwidth]{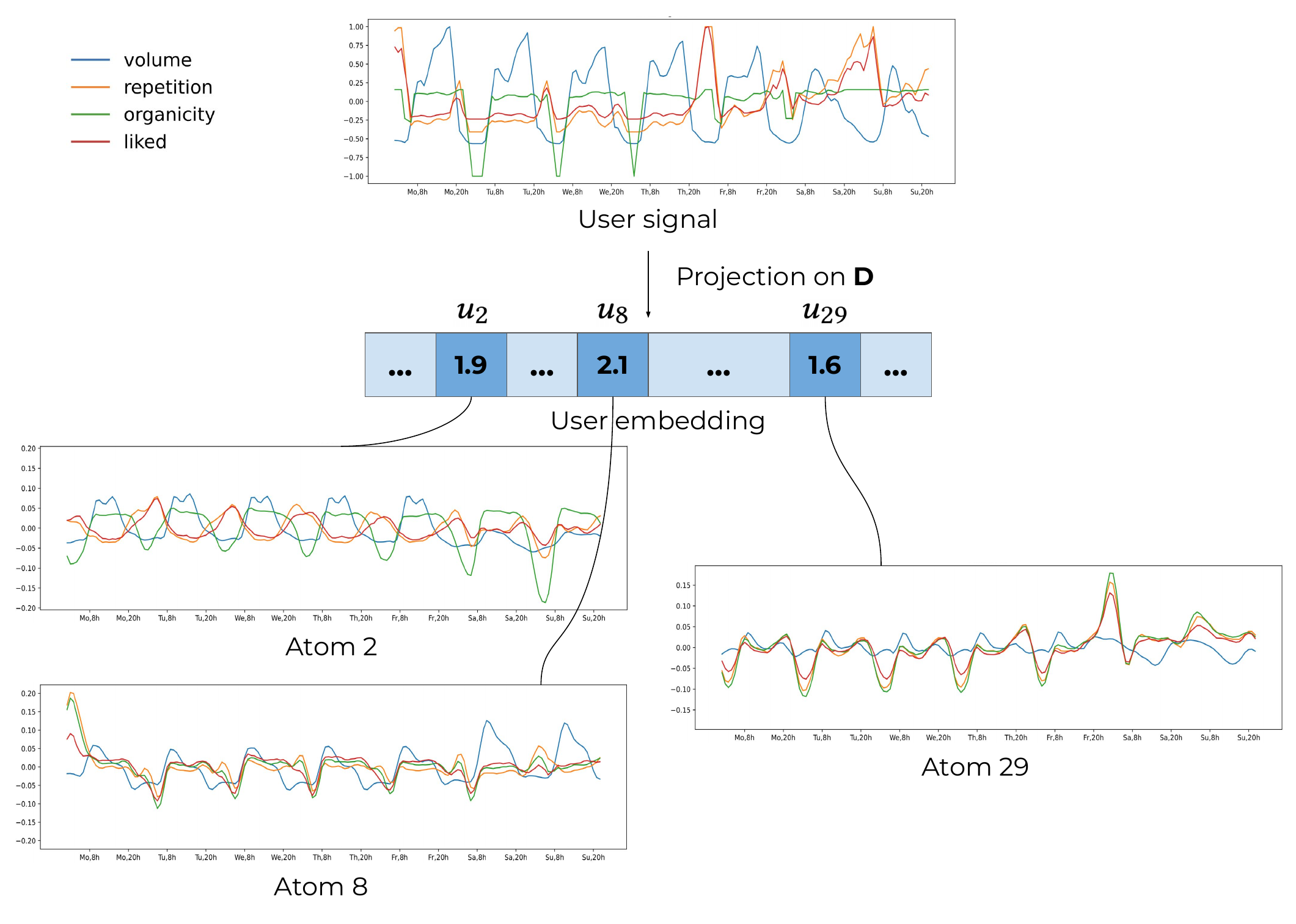}
  \label{fig:img2}
\end{figure}

While detecting patterns in $\textbf{D}$, the approximate resolution of this optimization program extracts user embeddings $\gamma$, which maps signals onto the dictionary under a sparsity constraint. 
With the chosen normalization, all signals are left to be in $[-1, 1]$ interval. We therefore expect the atoms learned on these signals to capture intra-user variations rather than general inter-user tendencies, at least more strongly than with a $L_2$ normalization.
By recovering $\gamma$ after the optimization process, we obtain an association of the users' consumption histories with the surfaced trends, thus we employ these associations as user embeddings, as depicted on Figure~\ref{fig:img2}. 
The sparsity constraint controls the $L_1$ norm of the embeddings, so a high $\lambda$ coefficient will tend to specialize certain atoms for the reconstruction of particular user signals.

The choice of the number of atoms, the number of iterations of the resolution algorithm, and $\lambda$ value can depend on the task one wants to focus on, for example a trade-off can be made between reconstruction scores, atom understandability, and a task-related score on a validation set.

\section{Framework evaluation}
Now that the PACE framework is in place, our attention turns to its evaluation in the context of activity-driven listening prediction. If the learned embeddings successfully capture recurrent consumption patterns, we can employ them to deduce the various recurring activities that influence listening sessions. We can do so as a result of a survey given to the Deezer users, asking them about the activities they engage in during listening. We pay particular attention to relating the results obtained to the \textit{a priori} regularity and frequency of the activities performed.

In this section, the exact number of atoms is set by practical considerations. Indeed, in order to keep a large enough panel of detected behaviors, without losing ourselves in too many patterns to analyze, we compute 32 atoms.
We implement both PACE framework and the following evaluation part in Python and make code available online\footnote{\href{https://github.com/deezer/modeling_activity_pace}{PACE source code : https://github.com/deezer/modeling\_activity\_pace}}, where the framework parameters are detailed.

\subsection{Dataset}
The dataset used is sourced from the research project RECORDS\footnote{\href{https://records.huma-num.fr/}{RECORDS web page : https://records.huma-num.fr/}}, funded by the French National Agency for Research, bringing together researchers from diverse backgrounds, studying practices on music streaming platforms.
Over a period of several weeks, surveys were sent to Deezer users by their email address, asking a wide range of questions about their cultural practices, musical tastes and socio-demographic variables. 

One of the asked questions was: ``In what contexts do you regularly listen to music?''. 
The users could then select multiple answers including:  \textit{Waking up} (\textbf{wake up}), \textit{When playing sports} (\textbf{sports}), \textit{On public transport (excluding driving)} (\textbf{transport.}), \textit{When working (including studying)} (\textbf{work}), \textit{Just before going to sleep} (\textbf{asleep}), and \textit{When receiving friends} (\textbf{friends}). Other activities are proposed, but we choose to focus on these 6 examples as they have a significant rate of positive answers, and are expected to be practiced with various regularities and frequencies, allowing us to test the effectiveness of our models.
Overall, there are around 10k respondents, which we reduce to 7k, removing the less active ones (less than 6 streams per day on average), for whom listening patterns are harder to detect.
Among the proposed answers, \textbf{wake up} and \textbf{asleep} are the less frequent activities (respectively 18\% and 15\%), while the others concern between one-third and one-half of the respondents (38\% for \textbf{transport.}, 39\% for \textbf{work}, 50\% for \textbf{sports}, and 47\% for \textbf{friends}). In comparison with the study carried out in \cite{Volokhin_Agichtein_2018}, respondents are far more frequent users of activity-driven listening.

Our evaluation task consists in predicting the answers of the survey from the user embeddings.
Thus, we frame the problem as 6 binary classifications, one per activity, each classification targeting whether the user has declared to practice the given activity or not. 
The implemented classifiers are logistic regressions models, optimized with 5 splits GridSearchCV. We evaluate the performance by the ROC AUC score on a test set ($33\%$ of the whole dataset, that is not involved in the learning of atoms).

We note that the pool of respondents is mainly made up of French users, with no significant gender bias, with the mean age being 32 years old (SD = 14). RECORDS researchers aim to publishing both anonymized listening histories and survey answers in open access during 2024.

\subsection{Classification Scores}
To the best of our knowledge, there are not many studies that attempt to predict activities from listening histories like ours. 
Therefore, we build several baselines to serve as complexity assessment and to position our model. 
The baseline models, based on the same modeling architecture but with training data capturing other types of information, are described hereafter: \textbf{(1)}~\textit{Total Volume}: from the logs, we retain the total number of streams performed during the covered period (1-dimension vectors); \textbf{(2)}~\textit{Gender \& age}: we encode the age group and the gender of the respondent (respectively coded in five and three categories, leading to 2-dimensions vectors); \textbf{(3)}~\textit{Other Activities}: for each user, we form a binary vector from the $6$ activity answers, then removing the activity we aim to predict (5-dimensions vectors).

Against those baselines, we evaluate our embeddings \textbf{(a)}, that we enrich concatenating vectors with \textit{Gender \& age} \textbf{(b)}.

\begin{table}[ht]
\caption{ROC AUC test scores of trained models.}
\begin{tabular}{lcccccc}
\toprule
  & \small{\textbf{wake up}} & \small{\textbf{transport}} & \small{\textbf{work}} & \small{\textbf{sports}} & \small{\textbf{friends}} & \small{\textbf{asleep}} \\
\midrule
\textbf{(1)} & 0.65 & 0.57 & 0.56 & 0.55 & 0.50 & 0.62 \\
\textbf{(2)} & 0.63 & 0.69 & 0.58 & 0.60 & 0.51 & 0.62 \\
\textbf{(3)} & \textbf{0.82} & \textbf{0.78} & \textbf{0.74} & \textbf{0.78} & \textbf{0.75} & \textbf{0.78} \\
\midrule
\textbf{(a)} & 0.69 & 0.67 & 0.63 & 0.56 & 0.61 & 0.73 \\
\textbf{(b)} & 0.70 & 0.71 & 0.65 & 0.60 & 0.60 & 0.74 \\
\bottomrule
\end{tabular}
\end{table}

As a reference, it's worth noting that a random variable following a \textit{well-informed} binomial distribution (with parameter $p$ equals to the proportion of positive responses for the label to predict) yields a score of $0.5$. Almost all the baselines are strictly above this value, showing their relevance to the task. However, in general, baselines \textit{Total Volume} and \textit{Gender \& age} show relatively low scores, while \textit{Other Activities} baseline is particularly strong. 
This high performance is not surprising, since there is some homogeneity in the user profiles that perform the same activities due to confounding factors such as age. 
The training data thus form an activity-driven listening profile, and represent information very similar to the target.
Also, whether derived from collected (\textit{Total Volume}), or declarative data (\textit{Gender \& age}, \textit{Other Activities}), the performance of baselines remains quite similar across all labels. 

In general, PACE shows intermediate performance between \textit{Total Volume} or \textit{Gender \& age}, and \textit{Other Activities}, which demonstrates that our embeddings capture a significant proportion of the information contained in the activity profile.
Contrary to baseline models, ours reveal greater disparities in performance among the labels. 
For instance, the most challenging label to predict from our embeddings is \textbf{sports}.
This weakness can be understood given the regularity focus of PACE, as sports activities are typically less regular and frequent compared to activities such as falling asleep (most people do it daily).

Incorporating the \textit{Genre \& Age} variables into the embeddings typically leads to improved performance compared to the two models taken separately. 
This demonstrates a certain complementarity between the sociological profile and the listening profile.
Specifically, the labels \textbf{transport.}, and \textbf{work} benefit the most from this association. 
This can be explained as these activities are closely associated with the working population in urban areas who commute to work daily. 
This user profile, which is likely prevalent in the dataset, may not be present in all age groups (e.g., among those under 18 who are still students, or retirees), emphasizing the importance of accounting for a sociological profile.

\subsection{Model Interpretation}

Logistic regression assigns coefficients to each label for every variable within the training vectors, i.e., each atom. The sign of these coefficients signifies whether the atom's impact is positive or negative, while their absolute values indicate the magnitude of their influence. Analyzing these coefficients enables us to gain insights into the predictive models. Specifically, linking these atoms to our understanding of preferred time slots for various activities can serve as a validity check.

\begin{figure}
  \centering
  \caption{Logistic Regression coefficients of the model based purely on PACE embeddings.}
  \includegraphics[width=.47\textwidth]{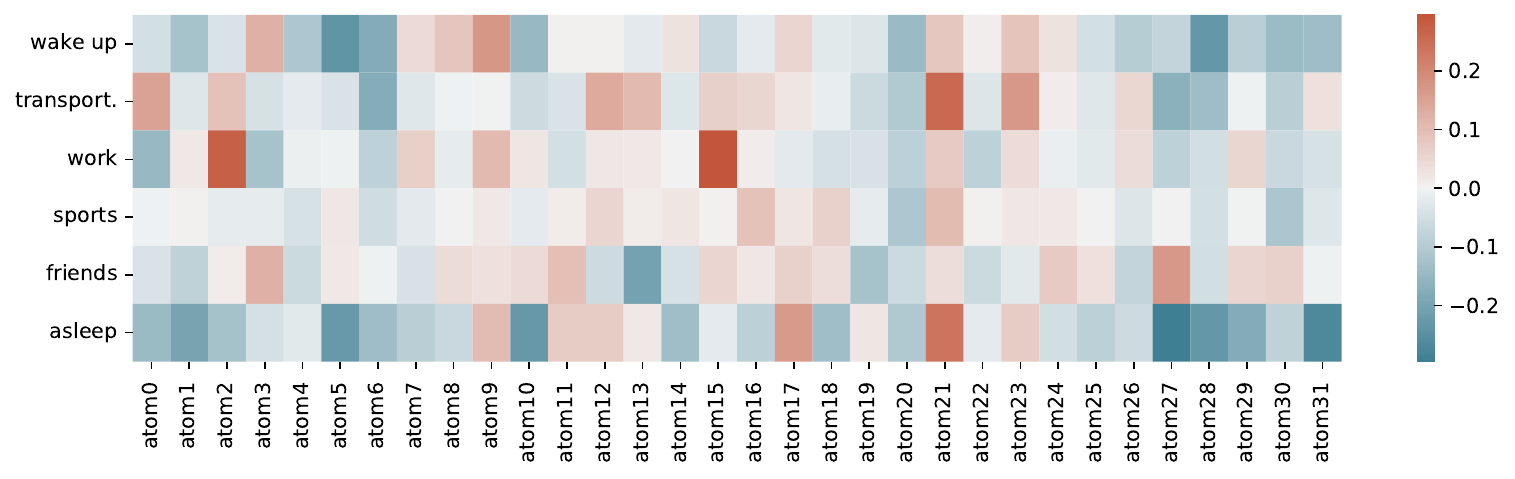}
  \label{fig:importance}
\end{figure}

\begin{figure}
  \centering
  \caption{Examples of detected listening patterns: atoms 0, 2 and 27.
}
  \includegraphics[width=.47\textwidth]{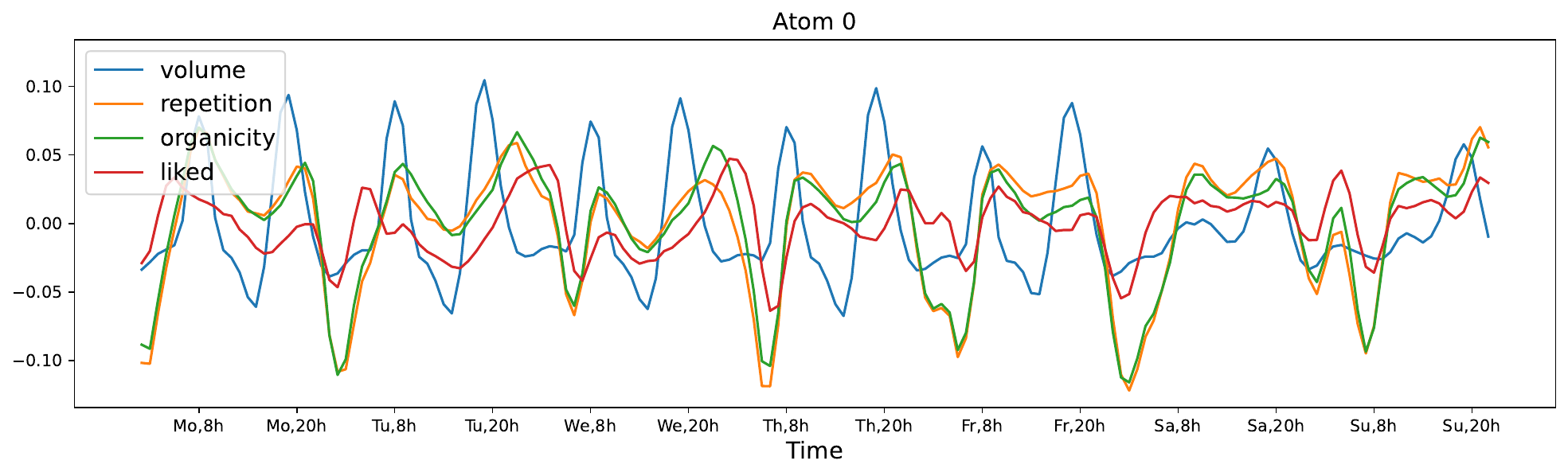}
  \includegraphics[width=.47\textwidth]{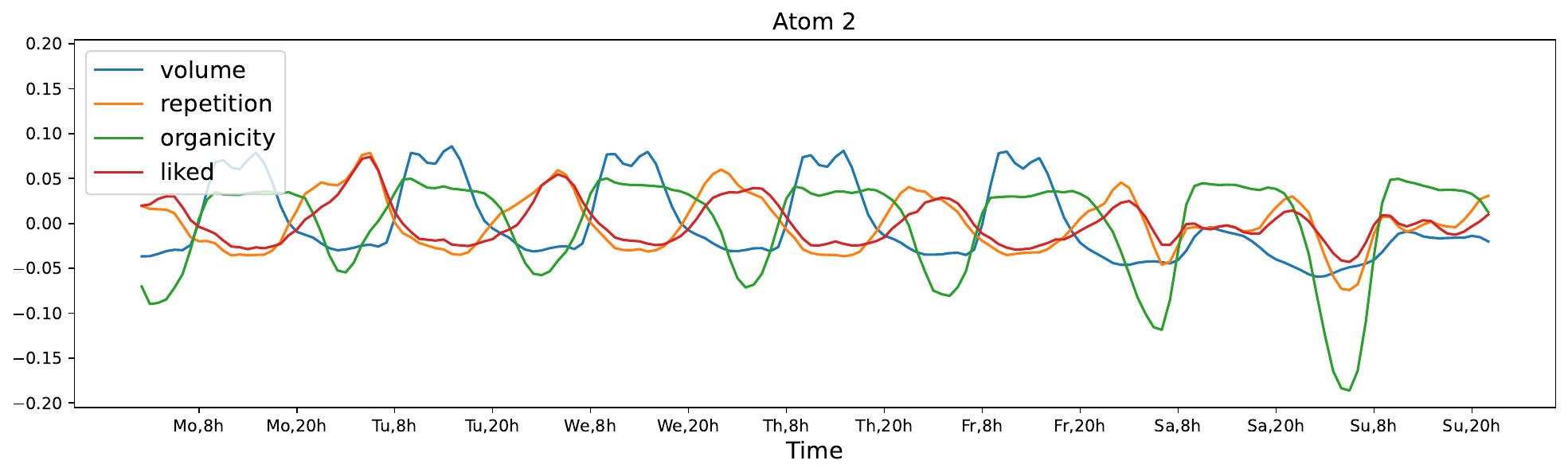}
  \includegraphics[width=.47\textwidth]{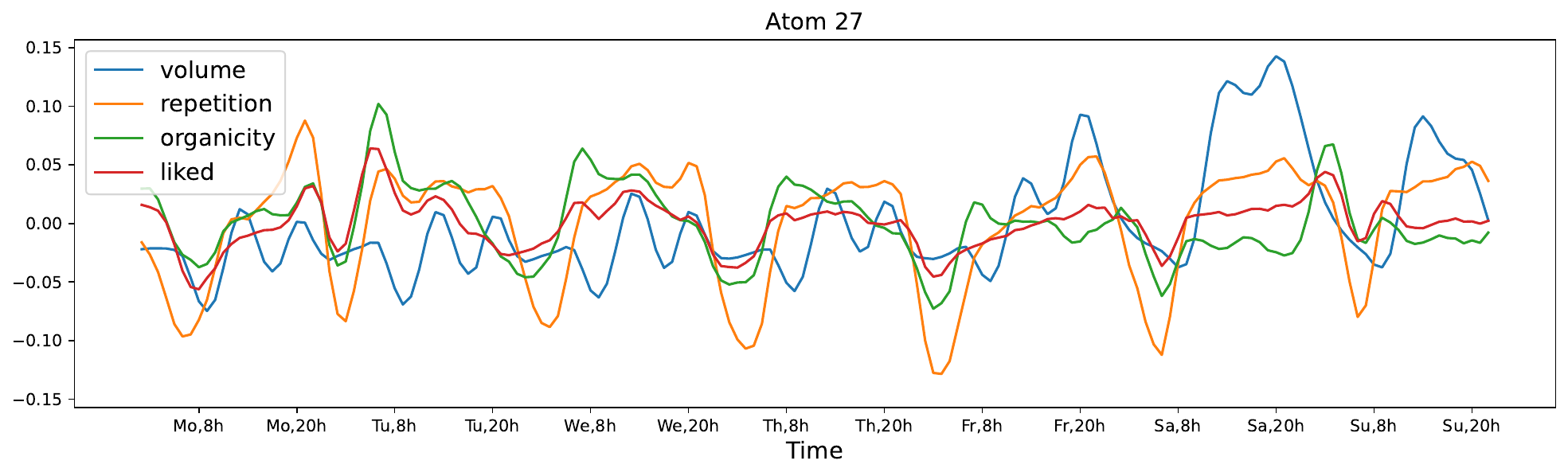}
  \label{fig:atoms}
\end{figure}

To do so, we can look for atoms having a positive coefficient for a single or few labels. 
Such atoms would identify the typical listening behaviors of the population performing the activity under consideration.
To this end, Figure \ref{fig:importance}  suggests matching, among others, \textit{atom $0$} for \textbf{transport.}, \textit{atom $2$} for \textbf{work}, and \textit{atom $27$} for \textbf{friends}. 
We display those atoms in Figure \ref{fig:atoms}.

\textit{Atom 0} shows peaks of activity on mornings and evenings of working days for \textbf{volume} channel, which seems consistent with listening while commuting. For \textit{atom 2}, the listening slots are concentrated on working days, and fit well with conventional working hours. 
In addition, there are small peaks in the mornings and evenings, suggesting also listening during commuting as in \textit{atom 0}, which pairs well with weekly workers' patterns. 
Furthermore, the \textbf{organicity} channel shows that listening at work is predominantly organic, whereas on weekend evenings, listening is much more algorithmic, revealing perhaps festive contexts with much less engagement. Lastly, \textit{atom 27} is dominated by channel \textbf{volume}, with a lot of streams during the weekend and in particular on Friday and Saturday evenings, which coincides with typical festive times with friends.

Generally speaking, in the example above, the atoms highlighted by the logistic regression coefficients are in line with our knowledge regarding the privileged times for practicing the different activities. We were also able to confirm the consistency of the highlighted time slots with the listening data of Deezer playlists dedicated to specific activities (partying, work, sport).
This suggests that PACE successfully achieves its purpose, but also shows great promise for picturing more precisely the behaviors involved in soundtracking by taking advantage of the several information channels.

\section{Conclusion}
In this paper, we introduced PACE, a user embedding method based on regularity in music consumption. With the PACE framework, users' consumption logs are represented by weekly time series, aggregating 4 signals, each representing a specific aspect of user behavior. With dictionary learning, understandable typical listening behaviors were extracted, forming a projection base of fixed size. The projections thus form user embeddings whose composition is easy to understand, since each coefficient represents the part taken by a particular atom in the reconstruction of the signal.

To validate our approach, we focus on activity-driven listening. 
%Here we posit that the regularity of the activities driving the listening will be reflected into the consumption data. 
Matching logs with declared activities allowed us to confirm the relevance of PACE, insofar as the most regular activities were the easiest to predict. 
Furthermore, studying the surfaced atoms and their relationship with the predicted labels enables us to unveil valuable insights about the nature of music consumption.

In \cite{yang2023nested} perspective, PACE takes advantage of MIR context, and proves its efficiency in predicting a situational context (activity). 
As a way to specialize embeddings for capturing regularity information, it would be relevant to study PACE in relation to other areas of musical listening practices, such as the diversity of content listened to, or appetite for new music.
For example, future work should investigate how the PACE embeddings could be integrated to recommender systems to improve contextual recommendation.

To progress on this task, a straightforward future work would be to integrate channels related to the listened content. This would facilitate the modeling of activities that are more connected to, say, a specific music genre. PACE is flexible enough to the possibility of integrating other information channels, which could provide new insights regarding regular listening behavior, but also be adapted to a multitude of downstream tasks where different information is needed and PACE might be of use.

\section*{Acknowledgements}
This paper has been partially realized in the framework of the “RECORDS” grant (ANR-2019-CE38-0013) funded by the ANR (French National Agency of Research)

\bibliographystyle{ACM-Reference-Format}
\bibliography{main}

\acrodef{CARS}[CARS]{Context-Aware Recommender Systems}

\end{document}